\documentclass[12pt]{iopart}

\usepackage{iopams}  

\usepackage{graphicx,color}
\usepackage{bm} 

\usepackage{amsthm}

\def\bra#1{\langle #1 |}
\def\ket#1{| #1 \rangle}

\def\e{\mathrm{e}}
\def\ii{\mathrm{i}}
\def\d{\mathrm{d}}

\def\cC{\mathfrak{C}}

\def\I{I}

\def\cU{{\mathcal{U}}}

\begin{document}
\title[Composing boundary conditions]{A dynamical composition law for boundary conditions}
\author{Manuel Asorey$^{1}$, Paolo Facchi$^{2,3}$, Giuseppe Marmo$^{4,5}$ and Saverio Pascazio$^{2,3}$}
\address{$^{1}$Departamento de F\'\i sica Te\'orica, Facultad de Ciencias, Universidad de Zaragoza, 50009 Zaragoza, Spain}
\address{$^{2}$Dipartimento di Fisica and MECENAS, Universit\`a di Bari, I-70125  Bari, Italy}
\address{$^{3}$INFN, Sezione di Bari, I-70126 Bari, Italy}
\address{$^{3}$Dipartimento di Fisica and MECENAS, Universit\`a di Bari, I-70126  Bari, Italy}
\address{$^{4}$Dipartimento di Scienze Fisiche and MECENAS, Universit\`a di Napoli ``Federico II", I-80126  Napoli, Italy}
\address{$^{5}$INFN, Sezione di Napoli, I-80126  Napoli, Italy}
\date{\today}

\begin{abstract}
We analyze the quantum dynamics of a non-relativistic particle moving in a bounded domain of physical space, when the boundary conditions are rapidly changed. In general, this yields new boundary conditions, via a dynamical composition law that is a very simple instance of superposition of different topologies. In all cases unitarity is preserved and the quick change of boundary conditions does not introduce any decoherence in the system. 
Among the emerging boundary conditions, the Dirichlet case (vanishing wave function at the boundary) plays the role of an attractor. Possible experimental implementations with superconducting quantum interference devices are explored and analyzed.
\end{abstract}


\vspace{2pc}
\maketitle

\section{Introduction} \label{sec-dpw}
The dynamics of a quantum particle in a bounded domain of physical space is a paradigmatic problem in quantum mechanics.
From the mathematical point of view, one must make sure that physical observables be properly defined in terms of self-adjoint operators \cite{von, QMbook}. This translates into suitable choices of boundary conditions, those most commonly used in physics being Dirichlet (vanishing wave function at the boundary), Neumann (vanishing normal derivative), and periodic ones. 
From a physical perspective the problem has a plethora of applications, ranging from atoms in cavities \cite{HarocheRaimond} to ions and atoms in magnetic traps \cite{atoms_ions}, to superconducting quantum interference devices (SQUID) \cite{reviewSQUID}.
The role and importance of boundary conditions has been recently stressed in an interesting article \cite{Wilczek}, where varying boundary conditions are viewed as a model of spacetime topology change. Notable applications arise in the context of the Casimir effect and its dynamical version, giving rise to photon generation in a microwave cavity with  time-dependent boundary conditions~\cite{photons}. 
In this Article we shall analyze the effect of a rapid alternating change of boundary conditions and show that the resulting dynamics yields novel boundary conditions. 
We shall make use of a general result \cite{aim}, that characterizes all possible unitarity-preserving boundary conditions.

\section{Trotter formula for alternating boundary conditions} \label{sec-trotter}

Let us consider a spinless particle of mass $m$ in a cavity $\Omega$ with a regular boundary $\partial\Omega$. We focus on the dynamics that arises when the cavity undergoes a rapid alternating change of boundary conditions with a time period $t/N$. Such an evolution is described by the following unitary operator
\begin{eqnarray}
\label{eq:2evol}
& & \underbrace{\left(\e^{-\ii t T_{U}/N} \e^{-\ii t T_{V}/N}\right)
\left(\e^{-\ii t T_{U}/N} \e^{-\ii t T_{V}/N}\right)
\dots
\left(\e^{-\ii t T_{U}/N} \e^{-\ii t T_{V}/N}\right)}_{N \; \mathrm{ times}} \nonumber \\
& & \quad =\left(\e^{-\ii t T_{U}/N} \e^{-\ii t T_{V}/N}\right)^N ,
\end{eqnarray}
where $T_U = p^2/2m = -\Delta_{U}/2m$ is the kinetic energy operator, with $\Delta_{U}$ a self-adjoint extension of the Laplacian $\Delta$, with given boundary conditions specified by the unitary
operator $U$ acting on the boundary Hilbert space $L^2(\partial\Omega)$ \cite{aim}. 
Notice that in general $\Delta_{U}$ and $\Delta_{V}$  do not commute. Physically, this corresponds to a rapid switching between two different boundary conditions, in the limit of frequent switching. 

Evolutions of this kind are familiar in the context of quantum chaos \cite{qchaos1,qchaos2} and in connection with the quantum Zeno effect \cite{exnerrev,7quests} (when the von Neumann measurements \cite{von} of the most familiar formulation \emph{\`a la} Misra and Sudarshan \cite{Misra} are replaced by frequent unitary pulses, as in the experiment by Itano \emph{et al} \cite{Itano}.) The analogy between these apparently unrelated dynamics has been explored by several authors \cite{qchaoszeno}.

The relevant question is to show whether in the $N\to \infty$ limit (when the time interval between the switches goes to zero, the number of switches goes to infinite, while the total time $2 t$ is kept constant),
the evolution is given by
\begin{equation}
\label{eq:evollim}
\left(\e^{-\ii t T_{U}/N} \e^{-\ii  t T_{V}/N}\right)^N \to
\e^{-\ii 2 t T_{W}},  
\end{equation}
in terms of a Hamiltonian $T_W$, with some boundary conditions $W$.

We will prove that this is the case, and in fact
\begin{equation}
\label{eq:2evollim}
W= U\star V = V\star U,
\end{equation}
where $\star$ is a commutative and associative product on the boundary unitary operators defined by
\begin{equation}
U \star V :=  \cC \left(\frac{\cC^{-1}(U) + \cC^{-1} (V) }{2}\right) .
\label{eq:stardef}
\end{equation}
Here $\cC$ is the Cayley transform that maps Hermitian into unitary matrices:
\begin{equation}
\cC (K) = 
\frac{\I  - \ii K}{ \I + \ii K }, \qquad 
\cC^{-1} (U) =  
-\ii \frac{\I-U}{\I + U}.
\label{eq:Cayley}
\end{equation}
Notice that the Cayley transform is not onto. Its range is the subset of unitary matrices whose eigenvalues are different from $-1$. Thus, strictly speaking, Eq.\ (\ref{eq:stardef}) has a meaning only for $U$ and $V$ in this subset. We will show what is the action of $\star$ on all boundary unitaries and prove that the eigenspaces with eigenvalues $-1$ are absorbing for the product. In particular, 
\begin{equation}
(-\I ) \star V = -\I,
\end{equation} for any unitary $V$.

The solution  makes use of  some mathematical results on  product formulae. In a seminal paper \cite{Trotter} Trotter proved that  
if $A$ and $B$ are self-adjoint and $C=A+B$ is self-adjoint on the intersection of their domains, $D(C)= D(A)\cap D(B)$, then the formula 
\begin{equation}
 \left( \e^{- \ii t A/N} \e^{-\ii t B/N}\right)^N   \to  \e^{-\ii t C}
\label{eq:Trotter}
\end{equation}
holds for $N\to \infty$.
This is the famous Trotter's product formula. 

Unfortunately, this formula cannot be applied to our case. Indeed, on the intersection of the domains of the Laplacians with different boundary conditions, $D = D(T_U)\cap D(T_V)$, the Laplacian is \emph{not} self-adjoint since the domain is too small, being defined by too many constraints (those of $U$ and those of $V$). Even more, it admits \emph{many} self-adjoint extensions, so that the meaning of $T_W$ in (\ref{eq:evollim}) is unclear: which boundary conditions $W$, if any, are obtained in the limit?

The answer is obtained by considering the quadratic forms associated to the operators (i.e.\ their expectation values), instead of the operators themselves. Notice, in fact, that the domain $D(t_U)$ of  the quadratic form 
of the kinetic energy,  $t_U(\psi) =  (\psi, T_U \psi)$ (see below), is larger than the operator domain $D(T_U)$, and there is a \emph{unique} self-adjoint kinetic operator $T_W$ associated to the quadratic form $(t_U + t_V)/2$ on the (dense) intersection $D = D(t_U) \cap D(t_V)$. The operator $T_W$ defined on $D(T_W) \subset D$ is called the form sum of $T_U$ and $T_V$ and is denoted by $T_W= (T_U \dot{+} T_V)/2$.

This idea, introduced by Kato \cite{Kato}, was elaborated by Lapidus \cite{Lapidus1,Lapidus} who found the ultimate version of Trotter's product formula~(\ref{eq:Trotter}):
If $A$ and $B$ are   self-adjoint and bounded below, and the intersection of their form domains $D(a)\cap D(b)$
is dense, then  Trotter's formula~(\ref{eq:Trotter}) holds when $N\to\infty$, with
\begin{equation}
C=A\dot{+}B,
\label{eq:Kato1}
\end{equation}
the form sum of $A$ and $B$.

As a technical remark notice that, as a consequence of the weakening of the hypotheses, the convergence of the product formula when the operator sum is not self-adjoint is in a weaker topology than in Trotter's case. More precisely, if $C=A+B$ is self-adjoint, then the convergence is  pointwise (in fact locally uniform) in $t$  in the strong operator topology. On the other hand, when $C=A\dot{+} B$ the limit is proved only on the average in $t$. Whether this convergence result is physically satisfactory or not, and whether it can be made stronger is a long-standing problem. For an interesting discussion see~\cite{ENZ}. 

\section{Boundary conditions for a free particle on an interval} \label{sec-interval}

In order to simplify our analysis and focus on concrete physical problems we shall restrict our attention to the problem of a particle moving in the interval $\Omega = [0,1]$ of the real line and set for convenience $m=1/2$. Apart from being the simplest mathematical case, the one-dimensional problem is also interesting in its own. Indeed, it can be implemented in a SQUID with a tunable junction, obtained by replacing the junction with an additional flux loop \cite{Vion,Cosmelli,Mooij}. On such devices different boundary conditions can be implemented by tuning the magnetic flux across the additional loop. More generally, a suitable combinations of tunable SQUIDs can be an experimental realization of the continuous interpolation among different topologies of the model introduced in \cite{Wilczek}.

The whole family of self-adjoint extensions of the Laplacian is in one to one correspondence with the possible boundary conditions given by the $2\times 2$ unitary matrices of $U\in \cU(2)$ \cite{aim}.
They are given by the Laplacian 
\begin{equation}
T_U  = - \frac{\d^2}{\d x^2}
\end{equation}
acting on the domain $D(T_U)$ of functions  belonging to $H^2(0,1)$,
the Sobolev space of square integrable functions $\psi$ with square integrable second derivative $\psi''$, and satisfying the  boundary conditions
\begin{equation}
\ii (\I +U) \varphi' = (\I-U) \varphi,  \qquad U\in\cU(2),
\label{eq:bc2}
\end{equation}
where
\begin{equation}
\varphi := \left(\begin{array}{r} \psi(0)  \\
\psi(1) 
\end{array}\right), \qquad \varphi' := \left(\begin{array}{r} 
- \psi'(0) \\
\psi'(1)\end{array}\right).
\label{eq:bc21}
\end{equation}

Physically, $U=-\I$ corresponds to Dirichlet boundary conditions, 
\begin{equation}
\psi(0)=\psi(1)=0, 
\label{eq:phiDir}
\end{equation}
while 
Neumann boundary conditions,
\begin{equation}
\psi'(0)=\psi'(1)=0,
\end{equation} 
are given by $U=\I$. The latter is a particular case of Robin's boundary conditions,  
\begin{equation}
\psi'(0)=-\tan \frac{\alpha}{2}\, \psi(0), \qquad
\psi'(1)=\tan \frac{\alpha}{2}\, \psi(1).
\end{equation}
given by $U= \e^{-\ii\alpha} \I$.
Moreover, 
\begin{equation}
U= \left(\begin{array}{cc}-1 & 0  \\ 0 & \e^{-\ii\alpha}  \end{array}\right)
\label{eq:U1end}
\end{equation} 
corresponds to Dirichlet at the left and Robin at the right,
\begin{equation}
\psi(0)=0, \qquad \psi'(1)= \tan \frac{\alpha}{2}\, \psi(1) . 
\label{eq:phi1end}
\end{equation}
For nondiagonal $U$ the boundary conditions at the two ends mix. For example,
pseudo-periodic boundary conditions, 
\begin{equation}
\psi(1)=\e^{\ii \alpha} \psi(0), \qquad \psi'(1)=\e^{\ii \alpha} \psi'(0), 
\label{eq:phiperiodic}
\end{equation}
are obtained  by 
\begin{equation}
U= \left(\begin{array}{cc}0 & \e^{-\ii\alpha}  \\ \e^{\ii\alpha} & 0 \end{array}\right) = 
\cos \alpha\, \sigma_x + \sin\alpha\, \sigma_y.
\label{eq:Uperiodic}
\end{equation}

In general, if the unitary $U$ has no $-1$ eigenvalues, the wavefunction $\psi$ can assume any values at the ends. Only the boundary values of its derivative are constrained is some way. These unitaries, corresponding to free ends, will be called \emph{regular}. 
Other case are not regular. For instance, one eigenvalue equal to $-1$, as in~(\ref{eq:U1end}) or (\ref{eq:Uperiodic}), corresponds to one constraint on the values of $\psi$ at the end, as in~(\ref{eq:phi1end}) or (\ref{eq:phiperiodic}),  respectively.
Finally, two $-1$ eigenvalues, i.e.\ $U=-I$, correspond to  two constraints on the wavefunction boundary  values~(\ref{eq:phiDir}).

\section{Quadratic forms} \label{sec-qforms}

Let $T_U$ be the kinetic energy operator. For any $\psi \in D(T_U) \subset H^2(0,1)$, an integration by parts gives
\begin{eqnarray}
t_U(\psi)&=& \left( \psi, T_U \psi \right) = -\int_{0}^{1} \bar{\psi}(x) \psi''(x) \d x 
\nonumber\\
&=& \int_{0}^{1} |\psi'(x)|^2  \d x - \bar\psi(1) \psi'(1) + \bar\psi(0) \psi'(0) 
\nonumber\\
&=& \| \psi' \|^2 - \langle \varphi | \varphi' \rangle.
\end{eqnarray}
We now try to rewrite the boundary form in a more convenient way by making use of the boundary conditions of $T_U$. In particular we want to trade the boundary values of the derivative for the boundary values of the function, in order to obtain a quadratic form
\begin{equation}
t_U(\psi) = \| \psi' \|^2 - \Gamma_U(\varphi), 
\label{eq:quadraticform}
\end{equation}
with $\Gamma_U(\varphi)$ a quadratic form of the boundary vector $\varphi$, given in~(\ref{eq:bc21}). 
In this way, the quadratic form (\ref{eq:quadraticform}) is defined on the Sobolev space $H^1 (0,1)$ of square integrable functions with square integrable \emph{first} derivative, at variance with the operator $T_U$ which required also the square integrability of the second derivative.
Notice indeed that, since 
$\psi'$ is in $L^2(0,1)$, 
then $\psi(x) = \int^x \psi'(y) \d y$ is a continuous function and  its boundary values $\varphi$ are well defined. On the other hand, the boundary values of its derivative $\varphi'$ would make no sense.

 We  distinguish among three possibilities according to the number of eigenvalues of $U$ equal to $-1$. Let 
\begin{equation}
\label{eq:specU}
U= u_1 \ket{\xi} \bra{\xi} + u_2 \ket{\xi^\perp} \bra{\xi^\perp},
\end{equation}
with $|u_1|=|u_2|=1$,  and $\bra{\xi} \xi^\perp\rangle = 0$,
its spectral decomposition (here $\langle \xi | \eta \rangle= \bar{\xi}_1 \eta_1 + \bar{\xi}_2 \eta_2$). We have:
\begin{enumerate}
\item
\label{case:K}
If  
$U$ is regular, i.e.\ $u_{1,2}\neq -1$, then $(\I+U)$ is invertible, and the boundary values of the derivative can be expressed in terms of the boundary values of the function
\begin{equation}
\varphi' =  K_U \varphi, 
\label{eq:Phi'Phi}
\end{equation}
where
$K_U = \cC^{-1} (U)$ is the inverse Cayley  transform of $U$, defined in~(\ref{eq:Cayley}).
Therefore, the boundary form $\Gamma_U$ is the expectation value of $K_U$, namely 
\begin{eqnarray}
\Gamma_U (\varphi) = \langle\varphi | K_U \varphi \rangle,
\end{eqnarray}
and $D(t_U) = H^1 (0,1)$, with free ends (no constraints on the boundary values $\varphi$).

\item
\label{case:-1}
If $-1$ is a nondegenerate eigenvalue of $U$, that is $u_1=-1$ and $u_2\neq -1$, 
then $U$ is not in the range of the Cayley transform~(\ref{eq:Cayley}), but from~(\ref{eq:bc2}) we get
$\langle \xi | \varphi \rangle = 0$, and
\begin{equation}
\langle \xi^\perp| \varphi' \rangle = 
-\ii \frac{1-u_2}{1+u_2}\,  \langle \xi^\perp| \varphi \rangle .
\end{equation}
Therefore, 
\begin{equation}
\bra{\xi} \varphi\rangle = 0, \qquad \Gamma_U(\varphi)= \ii \frac{1-u_2}{1+u_2}\,  |\langle \xi^\perp| \varphi \rangle|^2 .
\end{equation}

\item
\label{case:Dirichlet}
Finally, if $u_1=u_2=-1$, then $U=-\I$, and
\begin{equation}
\varphi = 0, \qquad \Gamma_{-I}(\varphi) = 0 .
\end{equation}
\end{enumerate}

\section{Composition law of boundary conditions} \label{sec-composition}

We now evaluate the limit of the alternating dynamics~(\ref{eq:evollim}).
According to~(\ref{eq:Kato1}) the product formula~(\ref{eq:evollim}) holds with the form sum
\begin{equation}
T_W = \frac{1}{2} \left(T_U \dot{+} T_V \right).
\end{equation}
Thus, the evaluation of the emergent  boundary condition $W$ in~(\ref{eq:evollim}) requires the computation of the sum
 \begin{equation}
t_W = \frac{1}{2} \left(t_U + t_V \right),
\end{equation}
 and its domain
\begin{equation}
D(t_W) = D(t_U) \cap D(t_V).
\end{equation} 

Again, we distinguish various cases according to  the number of eigenvalues $-1$:
\begin{enumerate}
\item
In the regular case,  when both $U$ and $V$  have no eigenvalues equal to $-1$, we get from~(\ref{eq:Phi'Phi}) that 
$K_W = (K_U + K_V)/2$, with no constraints on the wave-function boundary values $\varphi$. Therefore, one can write $W= U \star V$
with $U\star V$ given by~(\ref{eq:stardef}).
Explicitly, we get
\begin{equation}
W = U \star V =  \frac{\I-\frac{1}{2}\left(\frac{\I-U}{\I+U} + \frac{\I-V}{\I+V} \right)}
{\I+\frac{1}{2}\left(\frac{\I-U}{\I+U} + \frac{\I-V}{\I+V} \right)} .
\label{eq:regular}
\end{equation}

\item
If $-1$ is a nondegenerate eigenvalue of $U$, and $V$ is regular, then $D(t_W)= D(t_U)$, with the only constraint $\bra{\xi}\varphi\rangle = 0$. Therefore, the boundary forms $\Gamma_U$ and $\Gamma_V$ are nonzero and add up only on the orthogonal subspace, spanned by $\xi^\perp$. It is easy to see that
\begin{equation}
W= U\star V = - \ket{\xi} \bra{\xi} + w_2 \ket{\xi^\perp} \bra{\xi^\perp},
\label{eq:-1w2}
\end{equation} 
with
\begin{equation}
w_2 = \frac{1-\frac{1}{2}\left(\frac{1-u_2}{1+u_2} + \bra{\xi^\perp}\frac{\I-V}{\I+V} \xi^\perp\rangle\right)}
{1+\frac{1}{2}\left(\frac{1-u_2}{1+u_2} + \bra{\xi^\perp}\frac{\I-V}{\I+V} \xi^\perp\rangle\right)}.
\label{eq:w2}
\end{equation}

\item 
If  $-1$ is a nondegenerate eigenvalue of both $U$ and $V$, that is $u_1=v_1 = -1$ and $u_2, v_2 \neq -1$, then there are two possibilities
\begin{enumerate}
\item 
If the eigenvectors of $U$ and $V$ belonging to $-1$ are parallel,
that is $U$ commutes with $V$, then $D(t_W)=D(t_U)=D(t_V)$. Thus, the only constraint is $\bra{\xi}\varphi\rangle=0$
and $W$ has the previous form~(\ref{eq:-1w2}), where~(\ref{eq:w2})  particularizes into
\begin{equation}
w_2 = \frac{1-\frac{1}{2}\left(\frac{1-u_2}{1+u_2} + \frac{1-v_2}{1+v_2} \right)}
{1+\frac{1}{2}\left(\frac{1-u_2}{1+u_2} + \frac{1-v_2}{1+v_2} \right)} .
\end{equation}
\item 
If 
the eigenvectors $\xi$ of $U$ and $\eta$ of $V$ belonging to $-1$ are not parallel, then they span the whole space. The constraints $\bra{\xi}\varphi\rangle=0$ and $\bra{\eta}\varphi\rangle=0$ imply  Dirichlet's boundary conditions $\varphi=0$, so that
$D(t_W)=D(t_{-\I})$ and 
\begin{equation}
W= U\star V = -I .
\end{equation}

\end{enumerate}

\item
Finally, in the case $U=-I$ (or $V=-I$) then $D(t_W)= D(t_{-\I})$, so that $\varphi=0$ and
\begin{equation}
W = (-\I) \star U = U \star (-\I) =   -\I.
\end{equation}
\end{enumerate}

\section{Prediction: a SQUID circuit} \label{sec-previsione}

Summarizing, the limit boundary condition is $W=U\star V$, where $\star$ is given in terms of the Cayley transform~(\ref{eq:stardef}) for regular $U$ and $V$ (free ends $\varphi$), while all constraints on the wave-function boundary values are conserved by the product and inherited by $W$. 

Therefore, in one dimension, if $U$ and $W$ have independent constraints on $\varphi$, then Trotter yields Dirichlet boundary conditions, $\varphi=0$.
A particularly interesting case where this  happens is for pseudo-periodic boundary conditions, where $U$ and $V$ have the form~(\ref{eq:Uperiodic}) with two different phases $\alpha_1\neq\alpha_2$. In that case the boundary conditions read 
\begin{equation}
\psi(0)=\e^{\ii\alpha_1}\psi(1)=\e^{\ii\alpha_2}\psi(1), 
\end{equation}
and imply that  $\psi(0)=\psi(1)=0$
and $W=-\I$. 
This case can be experimentally implemented by means of a SQUID circuit where the properties of the Josephson junction are pulsed to mimic the Trotter evolution described above. The result (somewhat counterintuitive) will be a blockage of the electrical current through the circuit.

This situation is very different from the seemingly similar problem where the properties of the Josephson junction is not modified,  but a magnetic flux across the SQUID is pulsed in a similar way from $\alpha_1$ to $\alpha_2$. In this case the resulting evolution is very different because the boundary conditions are always the same 
but the Hamiltonians are not the Laplacian but have magnetic couplings
\begin{equation}
H_1= \left(-\ii\frac{\d}{\d x}+\alpha_1\right)^2,\quad H_2= -\left(-\ii\frac{\d}{\d x} +\alpha_2\right)^2.
\end{equation}
In this case  the Hamiltonian form sum  is
 \begin{equation}
H_3= \frac{H_1 \dot{+} H_2}{2} =  \left(-\ii \frac{\d}{\d x} + \alpha_3\right)^2,
\end{equation} 
where
\begin{equation}
\alpha_3=\frac{\alpha_1 +\alpha_2}{2}.
\end{equation}

\section{Conclusions} \label{sec-concl}

We have analyzed the quantum dynamics of a non-relativistic particle moving in a bounded domain of physical space, when the boundary conditions are rapidly changed. 
We have seen that the resulting boundary conditions is obtained though a dynamical composition law that can be viewed as a simple instance of superposition of different topologies.
Having in mind quantum gravity and the associated fluctuating topologies, one may speculate that according to our composition rule, the Dirichelet boundary conditions (that act as an attractor) play the same role as the classical paths in Feynman path integral, i.e.\ they completely dominate in the classical transition.

\ack 
This work was made possible thanks to the support of a INFN-MICINN Spanish-Italian grant.
This work of M.A.\ has been partially supported by the Spanish MICINN grants FPA2009-09638 and CPAN Consolider Project CDS2007-42 and DGIID-DGA (grant 2011-E24/2). G.M.\ would like to acnowledge the support provided
by the Santander\&UCIIIM Chair of Excellence Programme
2011-2012.

\end{document}